# Data-driven optimization of building layouts for energy efficiency

Andrew Sonta[1], Thomas R. Dougherty[1], Rishee K. Jain[1]

**ABSTRACT**

One of the primary driving factors in building energy performance is occupant behavioral dynamics. As a result, the layout of building occupant workstations is likely to influence energy consumption. In this paper, we introduce methods for relating lighting zone energy to zone-level occupant dynamics, simulating energy consumption of a lighting system based on this relationship, and optimizing the layout of buildings through the use of both a clustering-based approach and a genetic algorithm in order to reduce energy consumption. We find in a case study that nonhomogeneous behavior (i.e., high diversity) among occupant schedules positively correlates with the energy consumption of a highly controllable lighting system. We additionally find through data-driven simulation that the naïve clustering-based optimization and the genetic algorithm (which makes use of the energy simulation engine) produce layouts that reduce energy consumption by roughly 5% compared to the existing layout of a real office space comprised of 165 occupants. Overall, this study demonstrates the merits of utilizing low-cost dynamic design of existing building layouts as a means to reduce energy usage. Our work provides an additional path to reach our sustainable energy goals in the built environment through new non-capital-intensive interventions.

## 1. INTRODUCTION

The energy performance of buildings is largely driven by the operation of their energy-intensive systems. In commercial office buildings, the most energy-intensive systems are those that provide comfortable thermal and visual environments (i.e., heating, cooling, and lighting systems). The operation of these systems critically depends on the subjective experience of building occupants when they are using a building's spaces. As a result, there is no need to heat, cool, or light spaces that are not used at a particular point in time. These thermal and lighting systems are often controlled by zone, and in the case that even one occupant enters a zone, the systems must typically service the entire zone. This shared feature of building system operation contributes heavily to inefficiency (Yang et al. 2016), but it also creates the opportunity to optimize the design and management of building spaces and save energy through the individualization of building spaces.

Let us consider a hypothetical example of an office building with 4 teams, 4 members per team, and 4 shared rooms/zones. In this office, for the purposes of this example, one member of each team must be present in the office at any given time. All 4 teams have decided to operate on the following schedule: the first team member works from 12am–6am, the second from 6am–12pm, the third from 12pm–6pm, and the fourth from 6pm–12am. If the office is arranged such that each team occupies its own room/zone, there will be one person in each room at all times. In other words, all 4 rooms will have exactly 1 occupant in the room at all times. We refer to this situation as an example of high *occupant diversity* within each building zone, where diversity is a term used to describe the differences in activities or schedules. As a result of this diversity, the heating, cooling, and lighting systems for all four rooms will operate at all times. We could instead arrange the layout such that each occupant shares a room with their fellow shift-workers (e.g., all 12am-6am workers share a room). In this case, only one room, the occupied room, will need to be supplied with heating, cooling, and lighting throughout the day. The operation time of these systems would therefore be reduced by 75% compared to the first scenario.

---

[1] Urban Informatics Lab, Department of Civil and Environmental Engineering, Stanford University, 473 Via Ortega Rm 269B, Stanford, CA 94305

While this is an extreme example, the underlying dynamics apply to all buildings with shared spaces. In reality, the complexities and subtleties of occupant schedules, particularly in large office buildings, make it difficult to discern shared patterns of behavior. The decisions of creating building layouts, therefore, typically involve functional or hierarchical structures of organizations. In this paper, we investigate the possibility of optimizing layouts based on occupants' use of spaces.

Past research into building energy efficiency has focused on both building design as well as building operation. A large body of work has focused recently on the impact of occupant behavior on energy-intensive operation of building systems. Furthermore, researchers have investigated the role of design optimization, including through an occupant-centric lens, in reducing building energy consumption. Below, we discuss key findings from previous research and motivate the work presented in this paper.

## 1.1. Building design and energy efficiency

It is well known that the design of buildings has a large impact on their future operation, including energy efficiency (Lucon et al. 2014). When considering energy in building design, architects and engineers generally consider physical building parameters including orientation, materiality, fenestration, and choice of heating, cooling, and lighting systems (Basbagill et al. 2014; Petersen and Svendsen 2010; Roisin et al. 2008). The design process also considers, in addition to these physical characteristics of individual components, the impacts of layout on occupant behavior through the architectural programming process. This programming often is driven by intended building use (e.g., meeting rooms or workspaces in an office building). While this design process is typically regarded in terms of our subjective experience of the building, past research has shown that these choices for building layouts can influence the energy consumption of the building (Yang et al. 2016). For example, locating parts of a home that are more often used in the morning on the eastern side of the house could reduce the amount of heating, cooling, and lighting required for those spaces.

An advantage of considering the layout of the building in the context of energy efficiency is that layouts are often flexible, especially compared to other design considerations such as orientation and materiality. This is especially important due to the fact that our use of buildings evolves over time. Buildings originally intended for one purpose are often repurposed to suit changing needs, leading to different patterns of space utilization and rendering some physical design considerations obsolete. Moreover, as buildings lifespans are on the order of 40-100 years (Tanikawa and Hashimoto 2009), it is expected that the majority of energy consumption from the building sector will come from existing buildings rather than new construction for many upcoming years—years that are critical to our sustainable energy goals (Lucon et al. 2014). This importance of the existing building stock in addressing energy challenges suggests the need for new and innovative ways to reconsider the energy-intensive attributes of existing building design. New design methods that focus on building layouts are a promising means of addressing these challenges.

*1.2. The role of the occupant in energy efficiency*

As discussed above, building layouts are often driven by intended use. Research has shown that the actual use of spaces by building occupants, while extremely impactful to building energy consumption, is both very difficult to model and understand (Feng et al. 2015; Parys et al. 2009; Roetzel et al. 2014). For example, a seminal study showed that changes in occupant behavior can cause two-to-one discrepancies in actual versus expected energy consumption (Norford et al. 1994). Both occupant actions (e.g., interactions with building systems through thermostats, windows, etc.) as well as their passive use of space are important to energy consumption (Jia et al. 2017). This latter notion, passive space use, largely drives energy consumption in buildings that have systems that are able to respond (e.g., turn off or reduce service) based on occupancy information. These responsive systems are increasing in prevalence as energy codes and regulations are driving wider adoption of energy-efficient technologies in the building sector (Park et al. 2019).

*1.3. Optimizing buildings for energy efficiency*

As both design and occupant behavior are key factors that affect the energy efficiency of buildings, researchers have sought to leverage optimization of building design and operation as a tool to improve energy performance. These optimization approaches have generally focused either on building design pre-construction or operation of heating, cooling, and lighting systems post-construction. Design optimization is typically considered in the context of new construction (Basbagill et al. 2014). Such research has largely focused on physical building parameters in the early stages of design (Basbagill et al. 2014; Yu et al. 2015). However, because of the *coupled* effects of occupant behavior and building design on building performance, researchers have noted that occupant-related uncertainty can hinder confidence in early-stage design decisions supported by such optimizations (Hoes et al. 2011). These coupled dynamics have driven researchers to focus more on the operational phase of building—and particularly the importance of building occupants as a means of improving energy efficiency (Park et al. 2019).

Researchers have therefore developed data-driven tools that optimize control of energy-intensive systems in existing buildings, finding that optimizing control of systems through an occupant lens can enable large reductions in energy consumption, ranging from 15-70% depending on the types of systems analyzed (Agarwal et al. 2010; Balaji et al. 2013; Krioukov et al. 2011). Recently, researchers have found that an important component of such control optimization strategies is the explicit consideration of occupant comfort, which improves the subjective experience of the occupants but can reduce the energy savings possible (Ghahramani et al. 2018). This research shows the promise of introducing sustainable energy savings in existing buildings by controlling building systems optimally. However, this research also considers the dynamics of occupant behavior as given. To take full advantage of controllable building systems, there remains the opportunity not only to optimize how systems respond to building space use, but also to optimize how people themselves use spaces—an aspect we consider in this paper through the building layout.

The previous research discussed in this section has shown that we can optimize building designs before construction and optimize building controls after construction. However, the dynamic behavior of building occupants—behavior that enables research into building controls—can also enable new research into building design features that remain flexible once the building is erected. As discussed above, a focus on building layouts may offer a means for addressing energy gaps in design through a naturally occupant-focused lens. A key research gap, therefore, is the integration of design optimization with important characteristics of dynamic occupant behavior. While optimization of physical building parameters before construction will remain a promising area of inquiry, there is a need to be able to dynamically examine aspects of building design once its key operational parameter, the building occupant, enters the picture. Optimization of building layouts have been considered in terms of organizational structure and performance (Jo and Gero 1998; Lather et al. 2019; Lee et al. 2012), however, the direct optimization of building layouts to address the energy-efficient operation of building systems remains an area for continual research.

In this paper, our overarching research question is whether building layouts can be optimized to reduce energy consumption of energy-intensive systems by leveraging sensor data on occupant activities. We first describe our previously introduced methods for abstracting time series plug load data to states of occupant behavior. We then discuss our methods of analyzing zone-level diversity in occupant schedules, optimizing layouts, and simulating the impacts of adopting an optimized layout. We introduce a real-world case study, where we apply our methods to a floor of an office building with 165 occupants.

## 2. METHODOLOGY AND DATA COLLECTION

In this section, we describe our overall methodology for developing a framework that enables the optimization of building layouts for energy savings (Fig. 1). We first leverage ambient sensor data collected from plug load energy sensors at the individual desk level to describe occupants' use of space over time. We term this description the individual's *occupant schedule* (section 2.1). We then define a distance metric that can be used to describe the *zone diversity* in occupant schedules over several individuals (section 2.2). This distance metric creates the ability to cluster occupants spatially (i.e., create new layouts) in an effort to reduce this zone diversity and therefore building energy consumption (section 2.3). Due to the extremely high dimensionality of the possible solution space, we also introduce a genetic algorithm for creating occupant layouts based on expected energy consumption (section 2.4). For evaluation of our layout optimization algorithms, we introduce a data-driven surrogate model for simulating energy consumption of a building's lighting system based on occupant schedules (section 2.5). We describe the dataset used for analysis in section 2.6.

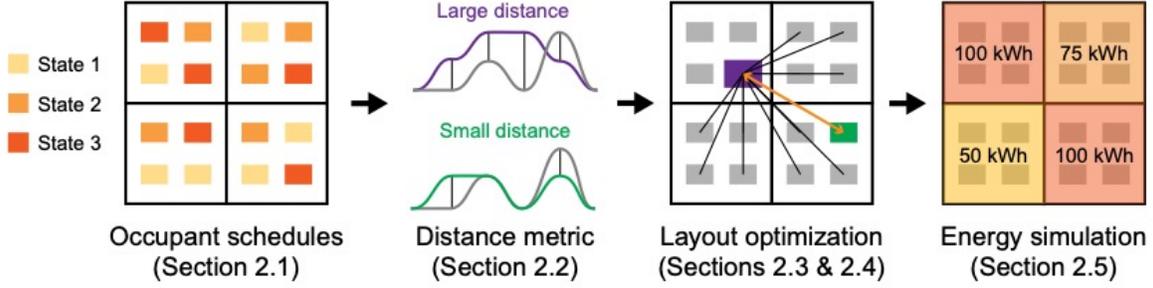

*Fig. 1: Outline of methodology.*

### 2.1. Data collection and preprocessing: learning occupant schedules from plug-load sensor data

Our analysis makes use of time series sensor data collected from plug load sensors installed at individual desks in commercial office buildings—data that provides insight into patterns of space use. We note that other sources of individualized sensor data that can describe individual patterns of space use could be adapted as the underlying data. We define the time series energy data as $X_{i,d}$ where i is the occupant index (for all occupants 1,...,I) and d is the day index (for all days 1,...,D). The total number of time steps is D x T, where D is the number of days and T is the number of time steps during the day (i.e., if we collect data at 15-minute intervals, T=96). Each entry in the data matrix, $X_{i,d}$ can be represented as a vector $\{x_1,...,x_T\}$. We leverage the method introduced in Sonta et al. (2018) to map the raw data to abstracted states of occupant activities: $X_{i,d} \rightarrow S_{i,d}$. This mapping is based on variational Bayesian inference with a Gaussian Mixture Model (VB-GMM) to cluster the time series data into discrete states. For each observation $x_t$, we introduce a latent variable, $z_i$ comprising a 1-of-K binary vector with elements $z_{t,k}$ for k=1,...,K, where K is the possible number of components that can be used to cluster the data. Given the set of weights for each component k, which we refer to as $\phi$, we write the conditional distribution of $Z$ as follows:

$$p(\mathbf{Z}|\boldsymbol{\phi}) = \prod_{t=1}^{T}\prod_{k=1}^{K} \phi_k^{z_{t,k}}$$

The conditional distribution of the observed plug load data, $X$, can therefore be written as follows, given the latent variables and component weights:

$$p(X|Z,\boldsymbol{\mu},\Lambda) = \prod_{t=1}^{T}\prod_{k=1}^{K} N(x_t|\mu_k, \Lambda_k^{-1})^{z_{tk}}$$

where $\boldsymbol{\mu}$ is the set of component means and $\Lambda$ is the set of component precisions (defined as the inverse of the standard deviations). Following standard Bayesian statistical practices, we introduce a Dirichlet distribution over the mixing coefficients and a Gaussian-Wishart prior over the mean and precision of each component. One of the key outputs of fitting this model is the number of components in $\phi$ that are non-zero. The resulting non-zero distributions are then used to cluster the data.

As discussed in Sonta et al. (2018), we use a two-step process for finding the number of components. If the initial clustering of the data results in two components, we separate out the higher-energy data and re-run the clustering algorithm. Our rationale for doing this is based on our domain knowledge of occupant behavior and plug load data—the higher energy data has high variability and is likely to represent multiple states of activity. Consistent with previous results, the model output is most commonly two components for the initial clustering and two components for the secondary clustering. We therefore apply this two-step two-component clustering to the data in this study. The result is a mapping **X → S,** where $s_{i,d}^t \in \{1,2,3\}$.

These three activity states, due to their construction, reflect intensity of energy use; we therefore refer to these states as low energy, medium energy, high energy. This intensity of energy use lends the physical interpretation of each activity state. For example, higher energy use values map to high energy activity states, which correspond to occupants actively using their workstations. Similarly, a medium energy activity state is likely to signify that some equipment has entered a power-saving mode without fully turning off. Past work has shown that this mapping of the ambient plug load data to occupant activity states constitutes an occupancy sensing strategy at least as accurate as other state-of-the-art sensing strategies, such as infrared sensors. An added advantage of the activity state strategy is that it offers additional information beyond presence/absence in that it describes the state of interaction with the workstation (e.g., a high energy activity state suggests full interaction with the workstation equipment). We refer to these individual time series of activity states as occupant schedules hereafter.

*2.2. Representing zone diversity of occupant schedules using Euclidean distance*

As discussed above, buildings provide energy-intensive services by zones, and the spatial efficiency of providing these services depends on occupant schedules. Therefore, a key question in understanding the operation of these systems, from the perspective of spatial efficiency, concerns the similarities or differences among the schedules of occupants within individual zones. We term these similarities and differences as the *diversity* in occupant schedules among occupants within a given space, and we operationalize this measure on time series data. Based on the work in Yang et al. (2016), we can define this diversity as the distance between the vectors describing time series schedules for each occupant in the zone. A range of distance metrics could be used, including cosine similarity, Manhattan distance, Euclidean distance, etc. Following previous research practice (Yang et al. 2016), we use the Euclidean distance for this study but note that the specific distance choice does not have a large impact on the analysis.

If our schedule data is structured as above $\mathbf{S}_{i,t}$, where i is the occupant index and t is the time index, we can compute Euclidean distances between the schedules for any two occupants. For example, the distance between occupant i and occupant j can be computed as follows:

$$d_{i,j} = \sqrt{\sum_{t=0}^{T} (\mathbf{S}_{i,t} - \mathbf{S}_{j,t})^2}$$

Using this distance metric, we can compute the distances between all occupants in a zone, forming a distance matrix. Normalizing this distance matrix by the total number of entries in the matrix (except the diagonal, since occupants' distance from themselves is 0), we have an average distance among all the occupant schedules within the zone, which we define as the *overall zone diversity*. With this metric, we can compare the diversity of occupant schedules for individual building zones to the actual energy consumption of the building systems. We would expect higher zone diversity to correlate with higher energy consumption, as was shown using physics-based simulation in Yang et al. (2016).

### 2.3. Optimizing layouts: Dimensionality reduction and occupant clustering

*2.3.1. Dimensionality reduction using Truncated Singular Value Decomposition*

Given occupant activity states and the notion of zone diversity in occupant dynamics, our next objective is to create optimal groupings of occupants in space—that is, to optimizing a building's layout as a means to minimize energy use. Time series sensing generally produces many signals over time for each sensor deployed. In our case, activity states are generally reported on the scale of 15 minutes, creating 96 signals per occupant per day, or up to 35,000 signals per year. Computation of distances between vectors of this size suffers from the well-documented curse of dimensionality, whereby distance functions lose their usefulness as the dimension of the space increases (Beyer et al. 1999). We therefore employ a common dimensionality reduction process known as Singular Value Decomposition (SVD), a generalization of Primary Component Analysis, as a means to reduce the dimensionality of our data. We note that the zone diversity metric introduced above can be computed either for the unreduced data or the reduced data—the definition holds for both perspectives.

A common technique in recommendation algorithms and dimensionality reduction, SVD allows the reduction in size of the data while still capturing valuable features (Leskovec et al. 2014). This dimensionality reduction is done by projecting the data with a set of orthogonal basis vectors representing the modes of variance in the system. These vectors are often referred to as the "concept space", as each vector represents some abstract concept which captures the variance. Truncating the lower variance orthogonal vectors before reconstruction yields a best approximation of the data in lower dimensional space, and we can select the number of dimensions depending on how much information we would like to retain during reconstruction. We apply SVD to a transposed version of our activity states: $\mathbf{M} = \mathbf{S}^T$ (where M has D x T time rows and I occupant columns) Here, the number of rows is expected to be much larger than the number of columns. The decomposition is as follows:

$$\mathbf{M} = \mathbf{U\Sigma V}^T$$

Here, $\mathbf{U}$ contains the eigenvectors of $\mathbf{MM}^T$, $\mathbf{V}$ contains the eigenvectors of $\mathbf{M}^T\mathbf{M}$, and $\mathbf{\Sigma}$ is a diagonal matrix containing the singular values of $\mathbf{M}$ ($\mathbf{\Sigma}^2$ contains the eigenvalues of $\mathbf{M}^T\mathbf{M}$). The shapes of these matrices are determined by $r = \text{rank}(\mathbf{M})$, where in our case, $r = I$. Therefore, $\mathbf{U}$ has the shape (D x T, I). These eigenvector matrices can be thought of as a rigid transformation in high dimensional space, which aligns the data according to the variance of the data. Thus the primary axis after the rotation will be aligned with the axis of highest variance, the second axis will be aligned with the second highest variance, etc. In practice, the matrices $\mathbf{U}$ and $\mathbf{V}$ map the data to a concept space. The concept space is defined by the shape of the primary eigenvectors in the system, which typically will provide some kind of intuition as to what is driving the variance. In occupant schedules, a concept space might identify the time at which a person usually arrives at the office to be a valuable indicator, or when they take their lunch break.

Our purpose for using SVD is to project very high dimensional occupant behavioral space to a much lower dimensional representation, permitting a richer and faster clustering process. After the matrix is decomposed into **U**, **V**, and **Σ**, the original data matrix **M** can be projected into concept space via a rigid rotation from **U**: $\mathbf{R} = \mathbf{U}^T\mathbf{M}$. The result of this projection is the condensing of the state data into an I x I matrix **R**, where each column, previously the length of the full time series, is now more densely represented, and the different columns correspond to the different occupants. Because **U** is orthogonal, the removal of the least powerful vectors prior to multiplication with **M** yields a projection of the data into the lower dimensional space with the least amount of lost value. For example, for a d-dimensional representation of the data, such that $d + d' = I$, the least significant d' eigenvectors/values in **U** are removed and the resulting representation of the full dataset **R'** will have the dimension d x I. This approximates the replacement of the smallest eigenvalues with a 0 term, which becomes redundant in the reconstruction and can be truncated without losing value.

*2.3.2. Stochastic constrained Expectation Maximization occupant clustering*

With this relatively low-dimensional representation of our activity data **S**, data-driven clustering of the occupants according to their activity states becomes more feasible. Here, we introduce a novel clustering algorithm based on the data representation **R'**. The objective of the clustering algorithm is to minimize the zone diversity metric from section 2.2 for the occupants within a building zone. Our problem setting has the real-world constraint that each of the building zones retain their same size at the end of the clustering routine to preserve the same overall occupant spatial density, preventing the use of standard clustering algorithms such as k-means. The intuition behind our novel approach is to minimize the zone diversity metric by spatially swapping occupants with other occupants that reduce zone diversity. By doing so, we can expect to reduce lighting consumption according to the relationship we have established between these metrics.

The mechanics of the algorithm are depicted in Fig. 2. First, we choose a random occupant/desk, which is associated with a building zone. We note that building zones do not all need to be the same size as depicted in Fig. 2. We then simulate a "swap" between this occupant and all occupants in the other zones of the building. The resulting swap will be the shift which had the greatest overall drop in zone diversity, which includes the null action of the node swapping with itself. We repeat this process until an iteration limit is reached.

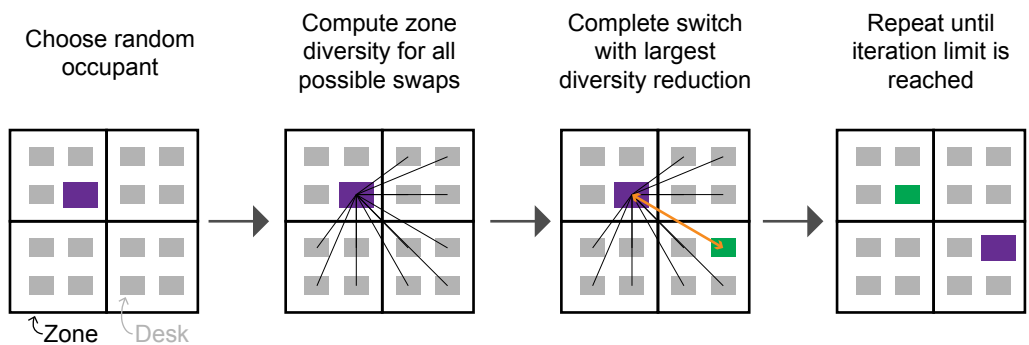

*Fig. 2: Occupant clustering algorithm.*

*2.4. Optimizing layouts: Genetic algorithm*

Our problem statement, to optimize building layout in order to reduce the energy consumption of building systems, has an extremely large solution space. If there are I occupants in a building and n possible zones to assign them to (of equal size m=I/n), then the number of possible assignments can be computed as follows:

$$\text{groups} = \frac{I!}{(m!)^n \times n!}$$

For example, with 50 occupants and 5 zones, the number of possible assignments is on the order of $10^{29}$. In addition to the large solution space associated with our problem, the effects of reassigning occupants are expected to be highly nonlinear. When optimizing in these circumstances, genetic algorithms have been shown to perform well (Nagy et al. 2017; Razavialavi and Abourizk 2017; Sonta and Jain 2020). We therefore implemented a custom genetic algorithm to assign occupants to desks and optimize building layouts, as described below.

Genetic algorithms belong to the class of evolutionary algorithms for optimization, originally inspired by the process of natural selection. The process begins by creating an initial set of design points—in our case, building layouts. Each building layout *x* in the initial population P is defined by the grouping of occupants to the zones of a building. A fitness function is used to evaluate the fitness of each design point *f(x)*. For this fitness function, we leverage a data-driven surrogate simulation engine that can be used to predict building energy consumption based on occupant schedules and other time series information, as discussed below in section 2.5. Once each design point is evaluated, a certain number of designs are selected to create a new generation of designs. In our case, we select the B best performing layouts, and, in order to maintain diversity in the population, we also select R random layouts. Among the best layouts and randomly selected layouts, two layouts are chosen at a time and recombined *c* times to form the layouts in the next generation. The first step is crossover, whereby for each desk location in each zone, the occupant selected to occupy that desk is a random selection of the two occupants in the original two designs. The next step is mutation, which occurs for each new individual with probability *m*. If mutation does occur, a random desk in each zone is swapped with a random desk from a random other zone. Crossover is meant to preserve the high-performing features that exist in the best-performing layouts in the previous generation; mutation is meant to introduce randomness so that the algorithm does not get stuck in a local minimum. Once a completely new generation is created from the previous generation, through crossover and mutation, the process repeats for *G* generations. The parameters, therefore, that must be chosen to run the genetic algorithm are the fitness function *f*, the population size |P|, the number of best performing layouts |B| and the number of randomly chosen layouts |R|, the number of new layouts *c* created for each chosen pair, the mutation probability *m*, and the number of generations *G*. Fig. 3 shows a visual representation of the algorithm.

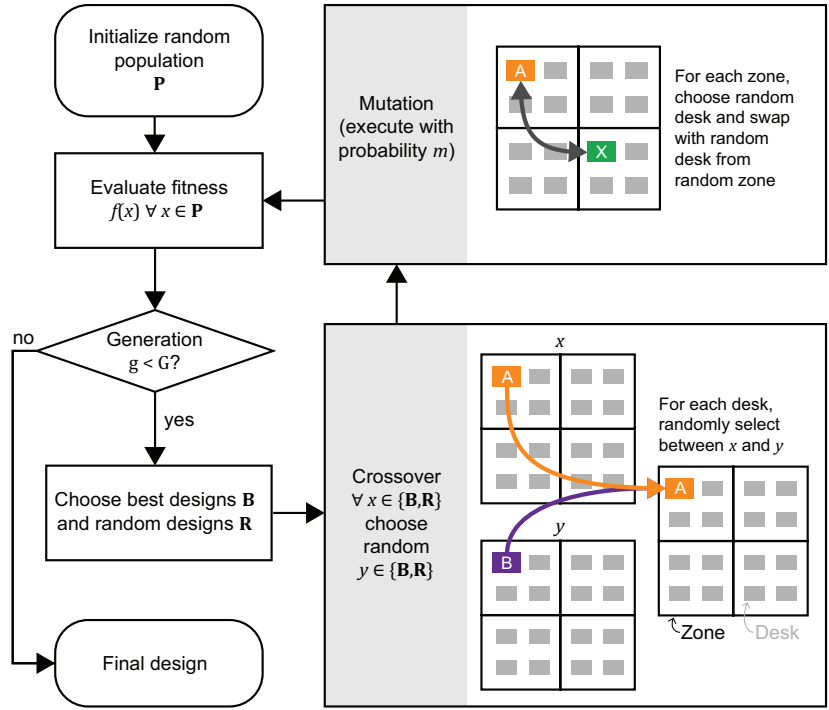

*Fig. 3: Genetic algorithm adapted for building layout optimization.*

## 2.5. Simulating lighting energy consumption based on building layouts

There are several viable simulation models for evaluating the expected energy consumption of different building layouts. These models fall into two main categories: physics-based thermodynamic models (e.g., EnergyPlus (U.S. Department of Energy n.d.)) and data-driven "surrogate" models (Aijazi and Glicksman 2016). Thermodynamic models have been shown to be particularly helpful when evaluating energy consumption from heating and cooling systems, though recently mixed models and data-driven surrogate models have become more prevalent. However, because our analysis focuses on lighting systems with direct control through occupancy sensors (as explained below), the thermodynamic models are more complex than necessary for this task. Additionally, the time required for their analysis is prohibitive for running our genetic algorithm optimization. Therefore, we sought to develop a data-driven surrogate model that utilizes machine learning to predict lighting energy consumption based on occupant schedules as well as standard time series features (e.g., time of day, day of week, etc.). We chose to test multiple linear regression (MLR), support vector regression (SVR), random forests (RF), and artificial neural networks (ANN) to determine the most robust surrogate model for our purpose.

The 7 specific features we use for prediction of energy consumption are as follows:
- $s_1$, $s_2$, $s_3$: number of occupants in each of the three energy states as defined above in section 2.1.
- Hour of day (0–23)
- Day of week (0–6)
- Weekday/weekend indicator (0 or 1)
- Zone number (0–number of zones)

The zone number is included to enable the model to adapt to zone-specific operational tendencies. For example, many lighting systems include daylight sensors, whereby the artificial lighting levels are lowered if enough daylight is present. This modulation would be expected to vary throughout the day for each zone, depending on orientation and other factors.

We implemented each potential surrogate model, described briefly here, using the scikit-learn package in Python (Pedregosa et al. 2011).
- *Multiple linear regression.* The simplest model, MLR seeks to predict energy consumption ($\hat{Y}$) as a linear combination of the model features (X): $\hat{Y} = \beta X + \varepsilon$, where $\beta$ is a vector of parameters and $\varepsilon$ is the error term. The fitting of the parameters involves minimization of the error term.
- *Support vector regression.* The model produced by SVR relies on a small subset of the training data known as support vectors. Errors within the bounds created by the support vectors (within margin $\varepsilon$) are ignored. Fitting an SVR model involves the following optimization:

$$\min \frac{1}{2}\|w\|^2 + C \sum_{i=1}^{n}(\zeta_i - \zeta_i^*)$$

  subject to the following constraints:

$$y_i - w^T \phi(x) - b \leq \varepsilon - \zeta_i$$
$$w^T \phi(x) + b - y_i \leq \varepsilon - \zeta_i^*$$
$$\zeta_i^{(*)} \geq 0$$

  where $w$ is the set of feature weights $\zeta_i$ and $\zeta_i^*$ are the residuals beyond $\varepsilon$ and $\varphi$ is a kernel function that is often nonlinear such as the Gaussian radial basis function. The SVR model has been previously applied to building energy prediction tasks with success (Jain et al. 2014).
- *Random forests.* The RF regression model is an extension of the decision tree model in regression form, in which the overall model aggregates (generally through averaging) the result from many independently fit trees. Each tree constitutes a series of decisions on the features (e.g., time of day is less than or greater than 6 am), and once the full series of decisions are made, a final value is chosen. RF models have successfully been applied to energy prediction in various settings (Ahmad et al. 2017; Wang et al. 2018).
- *Artificial neural network.* An ANN is an interconnected group of nodes, in which each node produces a signal according to the data it receives. ANN architecture generally involves an input layer (which receives the features), one or more hidden layers, and an output layer (which produces a prediction). These networks are fit using backpropagation. They have been widely used for energy prediction tasks (Ahmad et al. 2017; Ekici and Aksoy 2009).

*2.6. Empirical data*

We installed plug load energy sensors at each desk on a single floor or a large commercial office building in Redwood City, CA. The floor comprises 165 desks, of which 151 are occupied. The data collection period started August 1, 2019 and ended February 29, 2020. The sensors are Zooz SmartPlugs that communicate to a Samsung SmartThings hub through Z-Wave technology. The plug load sensors reported power consumption values any time the power consumption varied by more than 0.1 W. Consistent with previous work (Sonta et al. 2018), we aggregate the power consumption to 15-minute intervals, as modeling space use at this time frequency has been shown to limit noise while providing useful information in terms of building operation. As described above in section 2.1, we map the raw plug load data to energy states describing occupant schedules.

The office building is equipped with a lighting system that operates based on occupancy sensors, daylighting sensors, and schedules. The lighting zones are controlled by occupancy sensors across the building floor. If any lighting fixture within a zone senses motion over the past 20 minutes (10 minutes on weekends), all fixtures within that zone turn on. 11 of the lighting zones service all workspaces (the others service other small shared spaces such as meeting rooms, corridors, etc.) We restrict our analysis to the 11 zones that service workspaces, as we are interested in characterizing energy consumption in places where individual schedules can be modeled according to our data. The lighting energy data is available at 1-hour intervals for the full duration of the study.

Due to persistent sensor outages at the beginning of data collection, our analysis begins with data on October 1, 2019. In addition, due to a temporary shutdown of organizational activities over the New Year, we discarded data from December 16, 2019 to January 4, 2020. We therefore restrict our analysis to 132 full days of sensor and lighting energy data.

## 3. RESULTS

### 3.1. Increased zone diversity correlates with increased energy consumption

For each day over the data collection period, and for each of the 11 lighting zones in the building, we compute the zone diversity, as discussed above in section 2.2. We also compute the average energy consumption across lighting fixtures within a zone. We then complete a regression analysis for the relationship between zone diversity and energy consumption, as shown in Table 1. We find that for each zone, there is a positive relationship between zone diversity and energy consumption. We compute the t-statistic for the regression coefficient, and we find that the p-value for the t-statistic is significant at the 0.001 level for all zones. While the positive relationship is clear, both the slope of the relationship and strength of that relationship in terms of the $R^2$ value vary across zones. In Fig. 4, we show the data along with the regression lines and 95% confidence intervals for each zone. We note that this result serves as validation of the hypothesized relationship between occupant schedules and energy consumption. In addition, it serves as motivation for optimizing building layouts in order to reduce this diversity and therefore save energy.

*Table 1: Regression results.*

| Zone | Coefficient (Standard Deviation) | | t-statistic | p-value | $R^2$ |
|---|---|---|---|---|---|
| 0 | 422.5 | (78.29) | 5.396 | 0.000 | 0.165 |
| 1 | 492.4 | (52.64) | 9.354 | 0.000 | 0.354 |
| 2 | 828.2 | (106.7) | 7.762 | 0.000 | 0.288 |
| 3 | 532.4 | (80.00) | 6.655 | 0.000 | 0.288 |
| 4 | 527.9 | (35.33) | 14.944 | 0.000 | 0.627 |
| 5 | 749.6 | (126.3) | 5.934 | 0.000 | 0.253 |
| 6 | 514.3 | (40.78) | 12.611 | 0.000 | 0.288 |
| 7 | 989.2 | (51.83) | 19.087 | 0.000 | 0.666 |
| 8 | 177.6 | (39.70) | 4.472 | 0.000 | 0.117 |
| 9 | 486.9 | (20.94) | 23.256 | 0.000 | 0.757 |
| 10 | 97.1 | (60.90) | 4.878 | 0.000 | 0.157 |

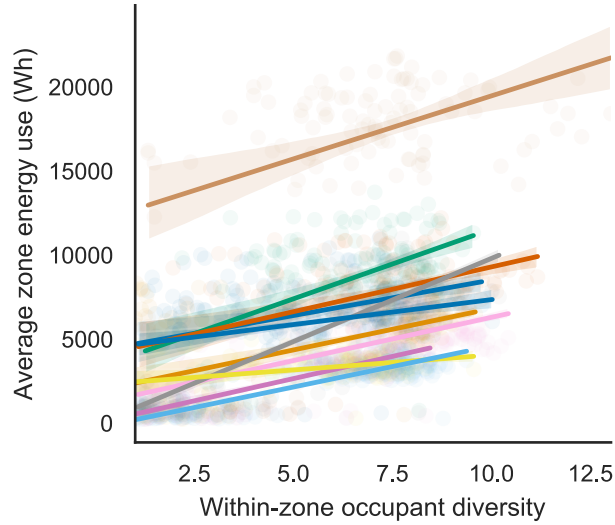

*Fig. 4: Relationship between zone diversity metric and energy consumption for each zone, along with regression fits and confidence intervals. Colors represent lighting zones.*

### *3.2. Lighting energy simulation is driven by occupant schedule data*

We tested the four models described above in section 2.5 by splitting the data into a training set and a test set, and then performing 5-fold cross validation on the training set for model development. The initial split into training and test sets preserves the time series attribute of the data: the first 80% of the data, in terms of time, are used as the training set and the second 20% are used for testing. We preserved the time series attribute of the data in order to test the ability of the models to predict future events and also to enable time series visualization of the performance of each model.

For the MLR, ANN, and SVR (non-tree-based) models, one-hot encoding is used for the day of week and zone number features, bringing the total number of features up to 23 features. Additionally, for these models, the hour of day feature is decomposed using sine and cosine transformations, to preserve the cyclical nature of the hour features (i.e., hour 23 is close to 0). Furthermore, the "state count" features (number of occupants in each energy state) are transformed using a sigmoid function. The intuition behind this transformation is that there are diminishing effects of having more than one occupant present in the zone (as shown in Fig. 5). This phenomenon occurs due to the fact that the lights only need to sense one occupant in order for all lights within a zone to turn on. Finally, the features are scaled to range between 0 and 1. Each of these steps were found to enhance the performance of the non-tree-based predictive models. These additional feature scaling steps are not used for the tree-based RF regressor, as the decisions are invariant to the scaling.

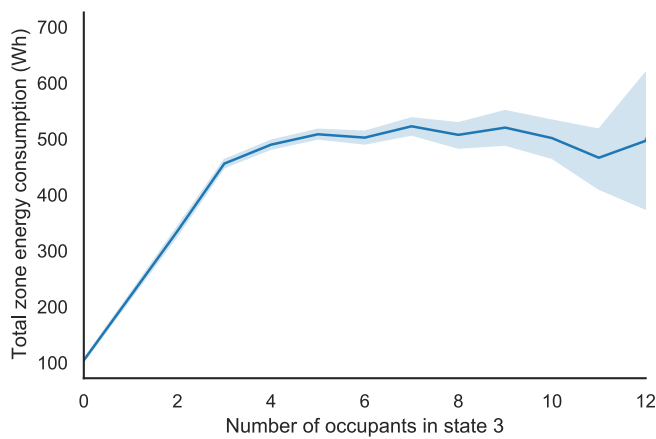

*Fig. 5: Total zone energy consumption vs. number of occupants in state 3.*

We find that, before tuning of hyperparameters, the ANN performs the best in terms of mean squared error (MSE) and explained variance ($R^2$), and the RF model performs the best in terms of mean absolute error (MAE) (Table 2). Because the squared error terms, MSE and $R^2$, exaggerate the importance of larger errors, we can infer that the RF model occasionally produces larger errors, while the ANN model produces a higher baseline of error values. While the ANN model takes the longest time to train, it requires less computation time to produce a prediction than the RF model. Based on these results, we chose the ANN and RF models for additional hyperparameter tuning.

*Table 2: Model results on 5-fold cross-validation.*

| Model | Mean Absolute Error (MAE) | Mean Squared Error (MSE) | Explained Variance ($R^2$) | Time for Training (s) | Time for Prediction (s) |
|---|---|---|---|---|---|
| Multiple Linear Regression | 9.55 | 141 | 0.534 | 0.0311 | 0.00198 |
| Support Vector Regression | 7.13 | 118 | 0.614 | 30.9 | 4.38 |
| Random Forest Regression | 6.11 | 98.2 | 0.678 | 2.82 | 0.0983 |
| Artificial Neural Network | 6.29 | 88.7 | 0.710 | 54.8 | 0.0105 |

Our hyperparameter tuning for both models also involved 5-fold cross-validation on the training set. For the ANN model, we tuned the hidden layer sizes, activation function, solver, and learning rate. We performed a grid search on these hyperparameters and found the best parameters to be a single hidden layer of size 100, the tanh activation function, the Adam solver, and a learning rate of 0.01.

For the RF model, we tuned the number of trees, minimum number of samples to produce a split in the tree, the minimum number of samples per leaf, the maximum depth of the tree, and whether the bootstrap methodology was used in model training. We found the best parameters to be 200 trees, minimum split size of 50, minimum samples per leaf of 2, maximum depth of 300, and use of bootstrap in model training.

We found that our tuned RF model outperformed our tuned ANN, with an MAE of 6.27, MSE of 87.1, and $R^2$ of 0.715 as estimated through cross validation. We also note that while hyperparameter tuning did improve the performance of each model, in both cases the improvement was very small. This high performance "out of the box" could enable wide adoption of similar surrogate models for the purposes of simulating energy consumption based on occupancy data.

Final results for the ANN and RF model, on both cross-validation and the test set, are shown in Table 3. Because the ultimate goal of our model is to be able to predict total energy consumption, as a result of features on the scale of 15 minutes, we also tested the performance of the ANN and RF models on a more aggregate level. We summed predicted and actual energy consumption to daily values and computed the $R^2$ metric on this aggregate level. As expected, both models benefit, in terms of their error rates, from this aggregation. However, the RF model benefits even further than the ANN.

*Table 3: Model results after hyperparameter tuning on both 5-fold cross-validation and final test set.*

| Model | Errors on CV | | | Errors on Test Set | |
|---|---|---|---|---|---|
| | **Mean Absolute Error (MAE)** | **Mean Squared Error (MSE)** | **Explained Variance ($R^2$)** | **Explained Variance ($R^2$) Hourly** | **Explained Variance ($R^2$) Daily** |
| Tuned Random Forest Regression | 6.27 | 87.1 | 0.715 | 0.740 | 0.834 |
| Tuned Artificial Neural Network | 6.28 | 88.5 | 0.710 | 0.734 | 0.817 |

An additional strength of the RF model is its interpretability. Each split in a tree involves a decision on one of the features, which do not need to be scaled or one-hot encoded. As a result, the importance of each feature can be easily quantified and visualized. For the final RF model, we find the most important features to be time of day, and number of occupants in state 3 (Fig. 6). This importance of the occupant feature underscores the notion that the lighting system operation is driven by occupant behavior, and therefore there exists opportunity to save energy by adapting the layout of the building to the behavior of occupants.

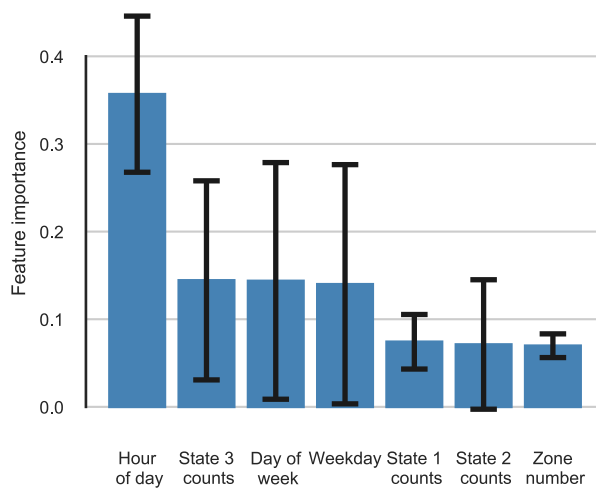

*Fig. 6: Feature importance for final random forest regression model.*

We note that the time for prediction using the RF model is roughly one order of magnitude above the time for prediction using the ANN model. In our case, these times were sufficiently small such that the difference between using them was inconsequential to our analysis. However, in situations with particularly high computing costs, the ANN model, while slightly less accurate, could be used to improve computational costs. That said, the RF model will always maintain a much higher level of interpretability.

### *3.3. Optimizing layouts reduces energy consumption according to simulations*

In this section, we show the results of our clustering-based and genetic algorithm optimization methods on two examples: a synthetic example that demonstrates the underlying mechanics of the optimization routines and the surrogate simulation model, as well as the empirical data collected from the office building in Redwood City.

#### *3.3.1. Synthetic example*

Here, we introduce a simple example of an office building based on the hypothetical example from the introduction, though updated to be more representative of common office behavior. Our purpose in showing this synthetic example is to demonstrate the mechanics of the optimization methods in creating new building layouts based on occupant activities. In this synthetic example, there are four different "archetypes" of occupant schedules. Each archetype defines the following behaviors:
- **Arrival.** Time when occupant arrives at the building, transitioning from low energy state to a medium or high energy state.
- **Lunch.** Time and length of occupant's lunch, in which the occupant transitions to a low energy state for the duration of lunch.

- **Meeting(s).** Time(s) and length(s) of the occupant's meeting(s), in which the occupant transitions to a low energy state for the duration of the meeting(s).
- **Departure.** Time when occupant leaves the building, transitioning to a low energy state.

When the occupant is in a normal working state, between arrival and departure but not during lunch or a meeting, the occupant randomly transitions between high and medium energy states. This behavior models normal occupant behavior of taking short breaks throughout the workday. The four specific occupant archetypes used in this synthetic example are shown in Table 4.

*Table 4: Occupant archetypes used in the synthetic example.*

| Archetype | Arrival | Lunch | Meetings | Departure |
|---|---|---|---|---|
| 1 | 9am | 12pm (1 hour) | 3pm (1 hour) | 5pm |
| 2 | 9am | None | None | 4 pm |
| 3 | 11am | 3pm (1 hour) | 3pm (1 hour) | 7 pm |
| 4 | 7am | 11pm (1 hour) | 1 pm (2 hours) | 5 pm |

The hypothetical building includes four rooms, with nine desks per room. Each occupant follows one of the four archetypes and there are exactly 9 of each archetype. We simulate energy consumption for one day, leveraging the random forest model discussed above in section 2.5. For simplicity, we assign each room to be the same zone number (0), which in the empirical example, has roughly the same size of 9 occupants. We also assume the day is a Monday. We test two different layouts to show the effects of layout on energy consumption: a random layout, and a layout in which all 9 occupants of each archetype share a room. We find, leveraging the lighting energy surrogate simulation model trained on real-world data, that the archetype-based layout creates a situation in which the lighting system is able to respond to occupants' use of the space (Fig. 7). The result is significant energy savings of 22%.

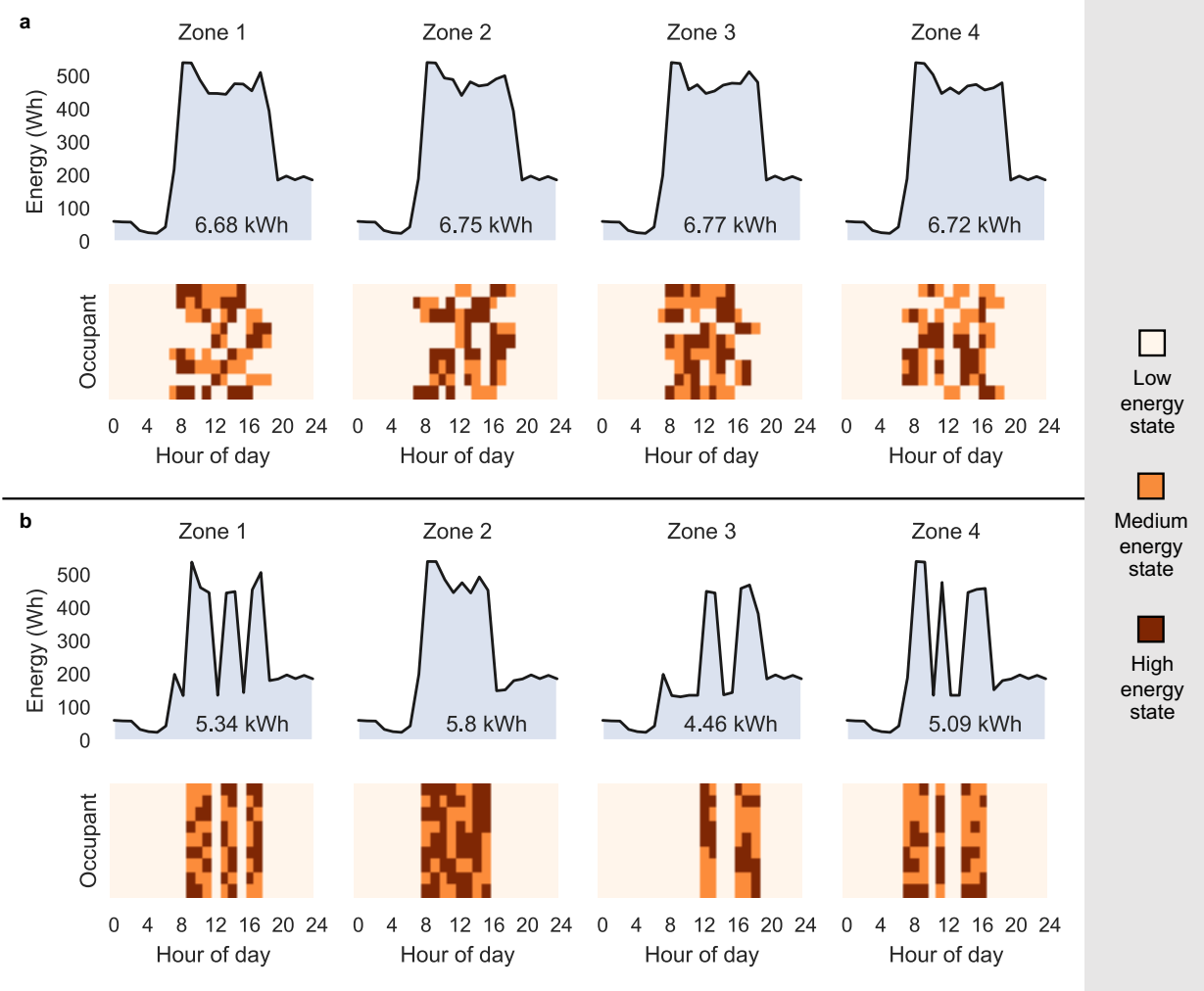

*Fig. 7: Synthetic example occupant schedules and simulated energy consumption for (a) random layout and (b) known optimal layout.*

We know the layout in which the building is arranged by archetype is optimal by inspection. This *a priori* knowledge of the system enables evaluation of the optimization routines introduced in sections 2.4 and 2.5. When we apply the clustering-based optimizer to this example, we find that the optimizer is able to arrive at the known optimal layout very quickly (Fig. 8). The genetic algorithm, on the other hand, quickly finds a near-optimal solution, but is unable to reach the fully optimal layout. This behavior of genetic algorithms—in which they come close to optimality but have final convergence issues—is well documented (Kochenderfer and Wheeler 2018). However, this synthetic example is relatively simple. In a real-world office building, with much more variation in occupant activities both in time and space, we expect the genetic algorithm to provide a sufficient solution as previous work has demonstrated that such methods can avoid local minima with sufficient variation in input data (Kochenderfer and Wheeler 2018).

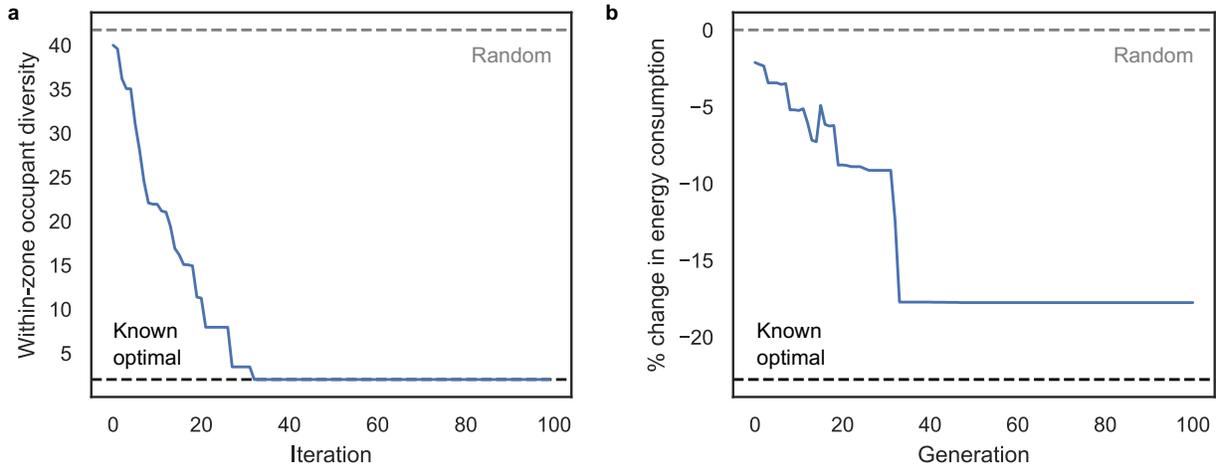

*Fig. 8: Optimization convergence on synthetic example, in relation to a random layout and the known optimal layout for (a) clustering-based optimization and (b) genetic algorithm. The same optimal layout produces the zone diversity value and energy values shown in the figure.*

*3.3.2. Empirical case study*

The synthetic example offers insight into the mechanics of the optimizers developed in this work, but the overly simplistic nature of that example limits claims that may be made about the extensibility of our framework. We therefore leverage the optimization algorithms and surrogate simulation model to optimize the layout of the Redwood City office building introduced in section 2.6. We simulate energy consumption using the full dataset of 132 days of occupant behavior. We start by estimating the energy consumption of the lighting system from 100 random building layouts as well as the existing building layout. We then apply the clustering-based optimization routine to the data with varying dimensionality (3, 5, 10, 30, 100, 151, and full dimensionality without reduction), as well as the genetic algorithm. We run each optimizer 100 times to create 100 layouts and ultimately a distribution of estimated energy consumption (Fig. 9). Overall, we find that the existing layout performs slightly better than the random layouts, but that further improvements on energy consumption can be realized by optimizing the layouts for energy efficiency. Additionally, increasing the number of dimensions used in the clustering optimizer improves performance.

We also computed the zone diversity in occupant schedules as introduced in section 2.2. Here, the zone diversity metric is computed for the entire 132 days, as opposed to daily in section 2.2. In general, we find that the zone diversity metric follows the predicted energy consumption (Fig. 10). However, we note that while the layouts produced through the genetic algorithm are among the best, the zone diversity metrics for the genetic algorithm layouts are substantially higher than for the clustering-based layouts.

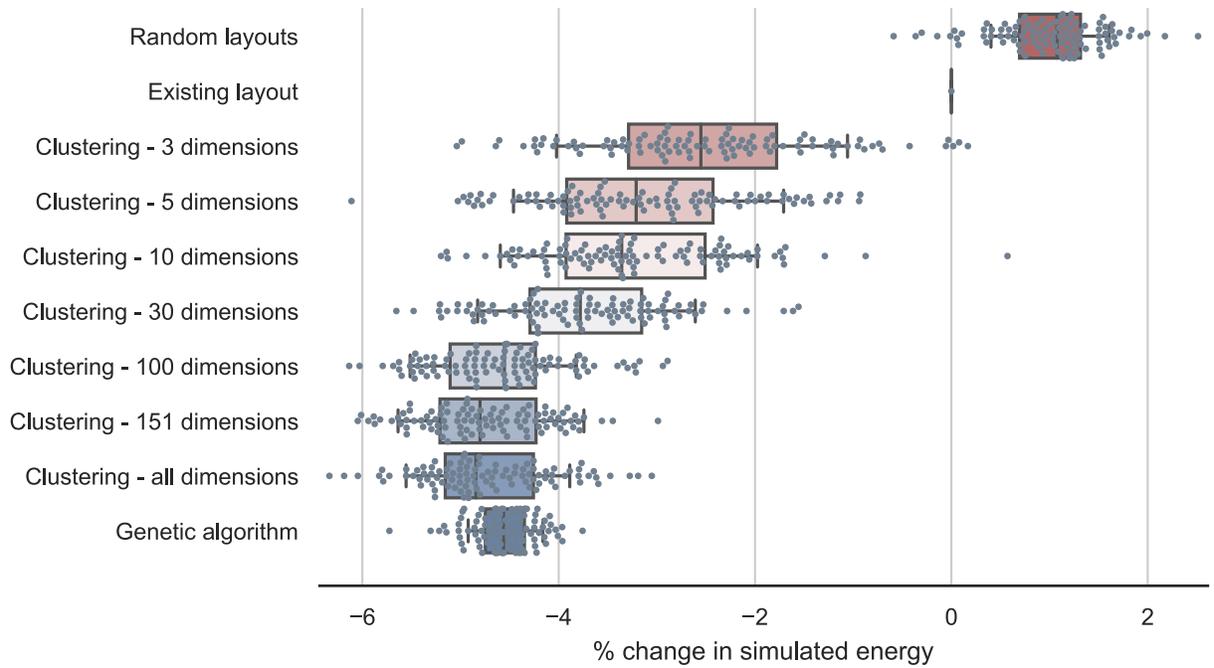

*Fig. 9: Simulated energy consumption (expressed as % change from the existing layout) for random and optimized building layouts.*

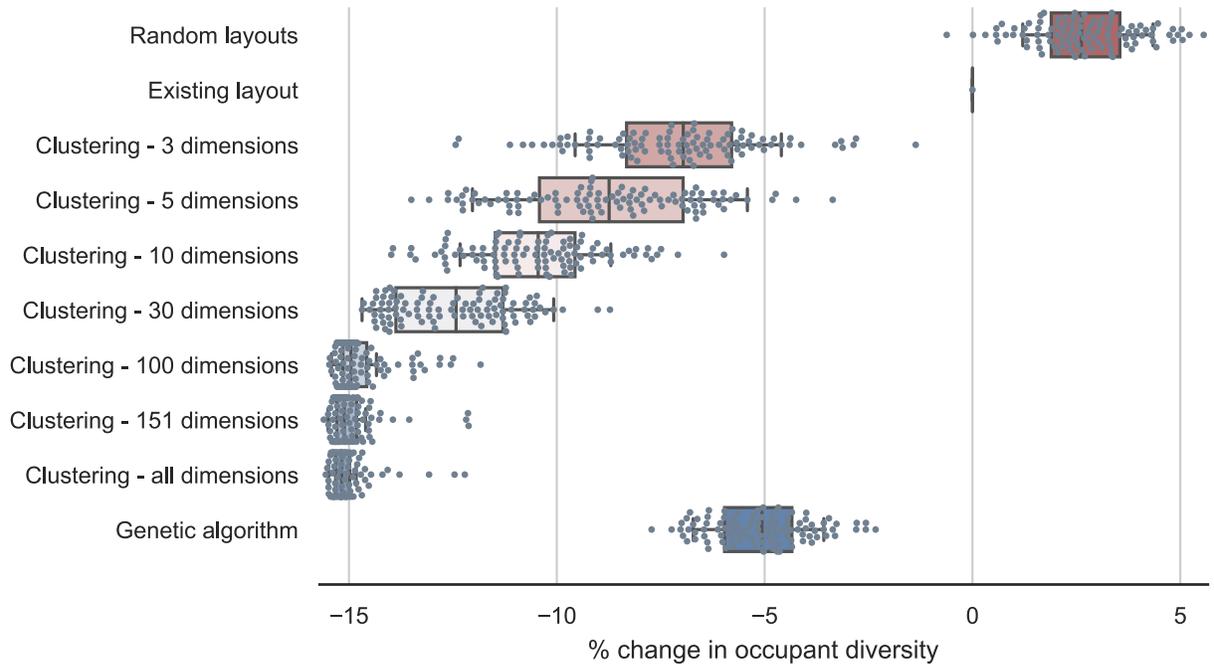

*Fig. 10: Change in zone diversity metric (expressed as % change from existing layout) for random and optimized building layouts.*

## 4. DISCUSSION

Our results indicate the possibility for significant energy savings of the lighting system through redistribution of occupants in space. This energy savings is possible by leveraging the occupancy-sensing feature of the lighting system. We note that while the heating, cooling, and ventilation system was not modeled in our case study due to data limitations, we would expect to find similar results as long as the thermal system is able to reduce service during periods without occupancy. The analysis of expected effects on these additional building systems is a rich area for further work.

While optimization creates reductions in energy consumption compared to the existing layout, it is interesting to note that the existing layout has a smaller simulated energy consumption than ~90% of the random layouts. This is not altogether surprising, as it indicates that behavior in the existing building is more aligned than if each occupant behaved independently. This result may suggest that people tend to adapt to the behavior of those around them, as has been documented in previous social science research (Chartrand and Bargh 1999). This question of how people adapt their behavior to their surroundings has potentially large implications for interpretations of our results. Underlying our analysis is the assumption that individual behavior will not change when occupant layouts are changed. We note that this assumption is a limitation of our model, and that behavior can be expected to change in some way in response to new desk assignment changes. The direction of the impact of these behavioral changes on energy consumption is unclear. However, our finding that the existing layout is more efficient than a random layout suggests that people adapt their behavior to match those around them. It is therefore possible that when we create layouts in order to reduce within-zone diversity in behavior, occupants may naturally choose to adapt their behavior and therefore further reduce within-zone diversity. On the other hand, it may be possible that occupants may choose to alter their behavior in other ways, either by attempting to maintain social behaviors that precipitated out of their previous setting, or perhaps by choosing to separate themselves from the others in their new surroundings. We note that the evidence from this paper seems to suggest that individuals, at least to some extent, tend to assimilate their behavior to those around them.

A key finding from this work is that the reduction in energy consumption from both the clustering-based optimization (when enough dimensions are used) and the genetic algorithm is roughly the same (Fig. 9). This is particularly interesting because the clustering-based optimization procedure does not include any information about the surrogate model used for predicting energy consumption, whereas the genetic algorithm uses that model as feedback during its execution. We argue that one can reasonably view the genetic algorithm as closer to a "best-case scenario" for predicted energy consumption, given its explicit use of simulation as a feedback mechanism during optimization. Our finding that clustering (once enough dimensions are used to represent the data) performs just as well demonstrates the strength of the naïve clustering-based approach. When only a small number of dimensions are used for clustering, the optimization does result in energy reduction, though the effect is smaller. It is therefore important to ensure that enough dimensions are used for representation of the occupancy data.

The effect of clustering on the within-zone occupant diversity metric is clear: substantial diversity can be reduced by clustering occupants according to their schedules. The genetic algorithm also reduces this zone diversity, but to a significantly smaller degree (Fig. 10). This finding suggests that there are other factors beyond zone diversity that are addressed during execution of the genetic algorithm. We unfortunately cannot interpret what these factors are, but they could involve unique aspects of the lighting system's operation. It is also possible that the genetic algorithm is optimizing for uncertainty in the random forest surrogate simulation model, which could be a limitation of that approach.

We found that the clustering-based optimization performs about as well as the genetic algorithm. The genetic algorithm is only executable when a simulation engine is available, which makes the clustering-based approach more practical in situations when only occupancy data are available. We note, however, that if a simulation model is available, the genetic algorithm can be seeded with the layouts obtained through clustering. In our case study, the clustering-based layouts form a distribution with regard to simulated energy consumption (Fig. 9). In a preliminary analysis, we found that seeding the genetic algorithm with 50 clustering-based layouts and 50 random layouts created new layouts that performed as well as the best layouts from the clustering (the layouts furthest to the left on the distribution in Fig. 9). Therefore, this ensemble approach may be useful in reducing uncertainty around the expected outcomes from either approach.

As discussed in section 1.3, optimization of building design and system control can create significant energy saving opportunities in buildings. For example, Krioukov et al. (2011) found that occupant-driven control of lighting systems can lead to 50-70% energy savings. These strategies are critical for making use of controllable building systems and achieving energy efficiency. In some ways, these strategies can be seen as the "low hanging fruit" of optimal building performance (e.g., turning the lights off when no occupants are present). The research we present is this paper takes these strategies one step further: our methodology is not just about optimizing building system operation using information about occupant dynamics, but also optimizing characteristics of those occupants—the layout of workstations—to take full advantage of building systems. The approach we introduce can be considered in tandem with the more traditional approaches of controlling energy-intensive building systems, and it has the added benefit of not requiring expensive upgrades to building automation systems in order to achieve additional energy savings. Furthermore, the approach we introduce can also be integrated with other characteristics of building design, such as thermal comfort or organizational success, both of which can be tied to building layout (Nagarathinam et al. 2018; Peponis et al. 2007).

The ultimate goal of the design optimization in this paper is to reduce energy consumption, a goal that is important for reaching our sustainable energy goals and reducing costs for organizations. However, it is essential to note that there are other goals for the success of the building and organization that should be considered as well. Chief among these goals is productivity—perhaps the ultimate purpose of commercial office buildings—which is difficult to define and varies among different organization. Numerous factors have been shown to influence productivity, including occupant thermal comfort (Keeling et al. 2012), organizational cohesion, and others. The benefit of the occupant-driven optimization-based design approach we introduce here is that there is a natural extension to include other objectives (such as those discussed above) through a multi-objective optimization. We specifically note that to address organizational cohesion and collaboration, the approach of space syntax analysis (Bafna 2003; Peponis et al. 2007) may yield viable optimization objectives. Other graph theoretical approaches that are designed to leverage organizational needs (Lather et al. 2019) could also complement the approach we introduce here. The integration of occupant-centric design for energy with design for organizational outcomes will be a rich area of future research investigations.

## 5. CONCLUSION

In this paper, we explored the relationship between occupant behavioral dynamics and energy consumption from energy-intensive building systems. We introduced (1) a zone diversity metric adapted from the literature for comparison with empirical building energy data, (2) a clustering-based building layout optimization methodology made possible through dimensionality reduction, (3) a novel genetic algorithm for building layout optimization, and (4) a data-driven surrogate simulation engine for predicting lighting energy consumption from occupancy data. In a case study, we found a significant relationship between building occupant zone diversity and actual lighting energy consumption. We also found that our layout optimization methods can be expected to reduce lighting energy consumption by about 5% from the existing layout and 6% from a random layout. Overall, we show that reconsidering the design of layouts in existing buildings has significant potential for realizing energy savings. Additionally, the approach of changing layouts to achieve energy efficiency also enables the simultaneous consideration of other factors influence by building layout, including organizational performance and thermal comfort. These methods, when integrated with the many objectives that drive building management, will be critical to ensuring dynamic and lasting energy savings in buildings.